\documentclass[twocolumn,showpacs,amsmath,amssymb,prb]{revtex4}
\usepackage{graphicx}
\usepackage{amssymb}
\usepackage{amsmath}

\newcommand{\etal}{\textit{et al.}}

\begin{document}
\preprint{Hall}
\title{Contribution of Disorder to the Hall Effect in Bi$_2$Sr$_2$CuO$_{6+\delta}$}

\author{L. Fruchter}
\author{H. Raffy}
\author{F. Bouquet}
\author{Z.Z. Li}
\affiliation{Laboratoire de Physique des Solides, C.N.R.S.
Universit\'{e} Paris-Sud, 91405 Orsay cedex, France}

\date{Received: date / Revised version: date}
%
\begin{abstract}

The in-plane resistivity and Hall coefficient have been measured for the  single-layer compound Bi$_2$Sr$_2$CuO$_{6+\delta}$ for the whole range of doping states. The deviation of the Hall coefficient, $R_H$, from a high-temperature linear behavior and the temperature dependence of the Hall angle are both only weakly dependent upon doping, contrasting with Bi$_2$Sr$_{2-x}$La$_x$CuO$_{6+\delta}$ and Bi$_2$Sr$_2$CaCuO$_{8+\delta}$. This is in contradiction with former proposals that the transverse transport detects the formation of incoherent Cooper pairs in the pseudogap state. Conversely, the analysis of the data using a phenomenological angular dependent scattering rate clearly allows to distinguish between underdoped and overdoped states, and we propose that the maximum in $R_H(T)$ simply arises due to the combination of a large 
T-independent
scattering rate and an anisotropic temperature dependent one. 

\end{abstract}


\maketitle

The superconducting single-layer compound of the Bi-based cuprate family, Bi$_2$Sr$_2$CuO$_{6+\delta}$ (Bi-2201), shows remarkable properties. Its maximum superconducting transition temperature, $T_c \simeq$ 20~K, is outstandingly low, as compared to the other single-plane compounds, Tl$_2$Ba$_2$CuO$_6$ ($T_c = $ 90 K) and HgBa$_2$CuO$_4$ ($T_c =  97$~K). A strong incommensurate crystallographic modulation\cite{zhang91} was proposed as the origin of this low transition temperature\cite{zzli2005}. The normal state of Bi-2201 presents contrasted results as a function of doping. In the underdoped (UD) regime, a large Nernst effect well above $T_c$ was observed, as well as an anomalously small vortex entropy below $T_c$\cite{capan2003} (this is similar to another single-layer compound, (La,Sr)$_2$CuO$_4$). Recently,  superconducting fluctuations were shown to be destroyed in Bi$_2$Sr$_2$CuO$_{6+\delta}$ for an unexpected small velocity of the fluctuating superfluid\cite{fruchter2004}. Both these results may be interpreted as evidence for non Gaussian fluctuations in a large interval above the transition temperature\cite{ussishkin2002,puica2003}. However, it was shown
that linear conductivity fluctuations in Bi-2201 are conventional and follow the universal behavior for a two-dimensional superconductor, whatever the doping state. In particular 
these fluctuations 
display no signature of the pseudogap phase; neither does non linear fluctuation conductivity\cite{sfar2005}. In this context, disorder, which leaves the linear fluctuation conductivity unchanged, is an appealing alternative to explain how large incoherent phase fluctuations may contribute to a decrease of the superconducting transition temperature well below its mean-field value\cite{rullier2006}. To pursue this idea further, it is interesting to systematically investigate the transverse transport properties for Bi-2201, the Hall constant, for the whole range of doping. Although the understanding of this quantity is notoriously difficult, we will show that the Hall data for Bi-2201 differs from those for other Bi-based cuprates, revealing in particular a strong 
T-independent
scattering rate that accounts for the 
maximum in the Hall coefficient.

\begin{figure}[b]
\includegraphics[width= \columnwidth]{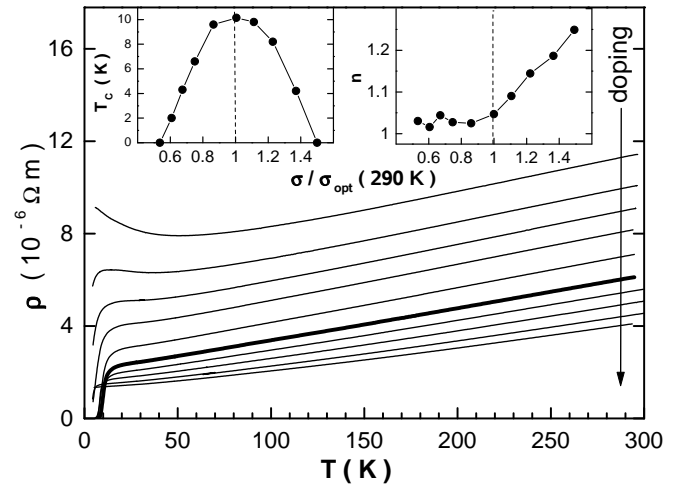}
\caption{Resistivity for different doping states. The arrow represents increasing doping (thick line is for the optimal doping). Insets shows the transition temperature and the exponent from a power-law fit above 150~K, $\rho(T)=a+b\,T^n$. The abscissa, $\sigma/\sigma_\text{opt}$(290~K) is an indication of the doping, varying in the range 0.1 - 0.35, according to 
Ref.~\onlinecite{konstantinovic2003}
.}\label{rau}
\end{figure}

\begin{figure}
\includegraphics[width= \columnwidth]{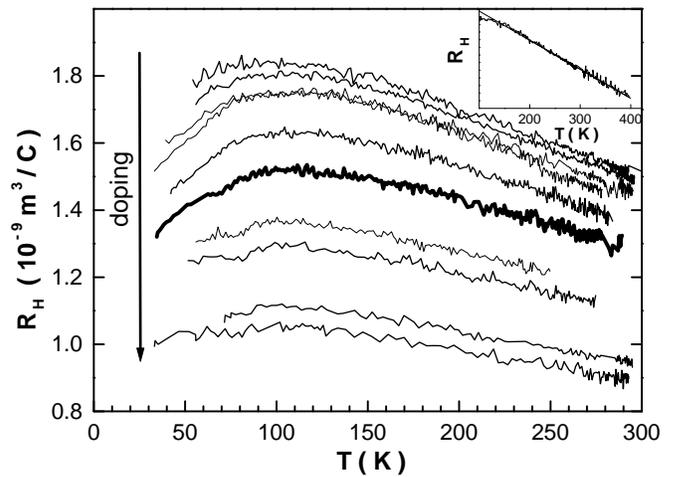}
\caption{Hall coefficient for  doping states as in Fig.~\ref{rau}. The thick line is for the optimally doped state.  Inset:  linear temperature behavior up to 400~K for the most underdoped state.}\label{RH}
\end{figure}

\begin{figure}
\includegraphics[width= \columnwidth]{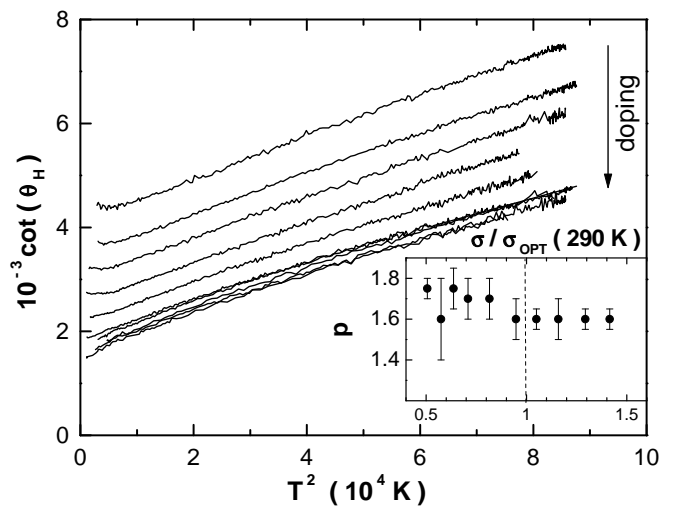}
\caption{$\cot(\theta_H)=\rho / R_H \, H$, at $H = 1$~T from the data in Fig.~\ref{rau} and \ref{RH}. The inset shows the exponent that best allows for a linear $\cot(\theta_H)(T^p)$.}\label{cotg}
\end{figure}

One 
single Bi-2201 thin film was studied for various doping states. It was epitaxially grown on a SrTiO$_3$ substrate and characterized by X-ray diffraction, as reported elsewhere\cite{li93}. First, annealing in pure oxygen flow at 420$^\circ$C led to the first maximally overdoped (OD) state. Successive lower doping states were obtained by annealing under vacuum at temperatures ranging from 220$^\circ$C to 280$^\circ$C.
The sample was mechanically patterned in the standard 6 contacts geometry. The Hall coefficient, $R_H = V_H t / I B$, where $V_H$ is the transverse Hall voltage,$I$ is the current flowing in the Hall bar and $t=2500$~\AA{} is the sample thickness, was obtained as a function of temperature during a single slow temperature drift 
(+~9$^\circ$C/hr).
The magnetic field was continuously varying in time, following a saw-tooth pattern with a maximum value of  0.5~T. The magnetic field dependence of $V_H$ allows to separate the Hall effect from the longitudinal resistivity contribution arising from the sample geometry imperfection (the Hall signal was typically 2\% of the total signal). Raw results for the longitudinal resistivity and the Hall coefficient are shown in figures~\ref{rau} and \ref{RH}. The  resistivity curves exhibit an upward curvature for the OD states and becomes linear near optimum doping and below, as evidenced by the power-law fits presented in Fig.~\ref{rau}. As repeatedly observed, the Hall coefficient (Fig.~\ref{RH}) exhibits a strong temperature dependence, which is one of the hallmark for the unconventional normal state of the cuprates. In agreement with results in Ref.~\onlinecite{rifi96}, a maximum in $R_H(T)$ is found in the range 90--100~K, almost independent from the doping state, which is somewhat higher than the 70~K value obtained in Ref.~\onlinecite{mackenzie92} for single crystals and for the slight maximum in Ref.~\onlinecite{bristol95}. The Hall coefficient does not show the commonly observed upward curvature at high temperature, but a linear temperature dependence which could be checked up to 400~K for the most UD state (Fig.~\ref{RH}, inset). It has been proposed that the cotangent of the Hall angle ($\theta_H$) may be more significant than the Hall coefficient. Indeed, within a scenario of spin charge separation for a Luttinger liquid, $\cot(\theta_H)$ is related to the magnetic scattering time alone, whereas $R_H$ is related to both the magnetic and the conventional longitudinal scattering time \cite{anderson91}. This scenario accounts for the experimentaly observed\cite{chien91} linear  $\cot(\theta_H)(T^2)$. As may be seen in Fig.~\ref{cotg}, $\cot(\theta_H)$ is clearly sublinear in $T^2$. It is actually better described as $\cot(\theta_H) = A + B\,T^p$, where $p \approx 1.6$ and possibly slightly increasing  with underdoping (Fig.~\ref{cotg}, inset). Such a failure was noticed by Ando \etal\cite{ando99} in the case of La-doped single crystals, Bi$_2$Sr$_{2-x}$La$_x$CuO$_{6}$: the exponent ranges from $p=1.85$ for the most UD (i.e. the largest La content) to $p=1.6$ for the most OD sample  (with $p=1.7$ at optimal doping). According to Ando \etal, this suggests, within the spin charge separation scenario, that the role of the spin degree of freedom is weakened, which would also explain the low maximum transition temperature (35~K)  as compared to other single-layer compounds. This is also in agreement with the results obtained on Bi$_2$Sr$_2$CaCu$_2$O$_{8+\delta}$\cite{konstantinovic2000}. In the present case, the reduction of the exponent is obtained in the whole range of doping, and it is noticeable that this occurs for a compound where the maximum $T_c$ is depressed to a lower value ($T_c \approx 10$~K).

\begin{figure}
\includegraphics[width= \columnwidth]{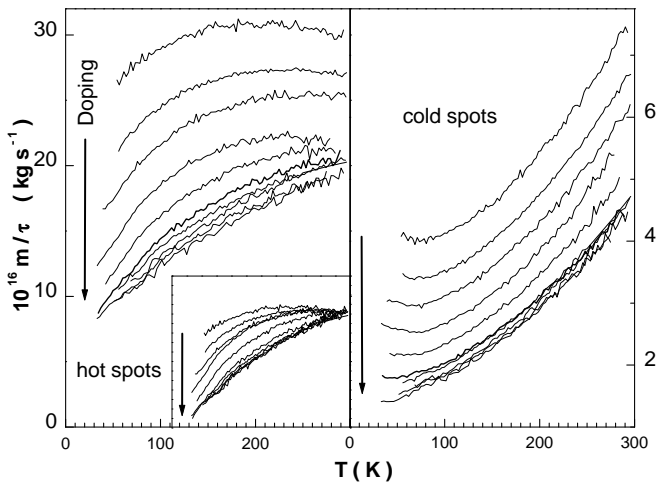}
\caption{Scattering time obtained from the phenomenological anisotropic mean free path model (see text). The thick line is for the optimal doping. Inset: $m_{hot}/\tau_{hot}$ normalized at 300~K, showing the superposition of the scattering rate for all overdoped states on a non-saturating curve.}\label{scattering}
\end{figure}

There have been many other theoretical proposals to account for the anomalous normal state properties of the cuprates (for a review, see Ref.~\onlinecite{stojkovic97}). A particular class is the Fermi-liquid model one, where an anisotropic scattering, as found in ref.~\onlinecite{hussey1996}, is introduced along the Fermi surface. The origin of this anisotropy is a much debated controversy; however a simple phenomenological model can be used to describe the transport data.
We have used the work of Ref.~\onlinecite{stojkovic97}, where a phenomenological description of the variation in the mean free path along the Fermi surface  was used (using a simple four-fold function): parts of the Fermi surface (`hot spots') experience a stronger scattering  than the rest of the Fermi surface (`cold spots'). The Hall coefficient and Hall angle can then be written:
\begin{equation}
R_H \approx R_H^\infty \, \frac{1+r}{2\sqrt{r}} \ ,
\text{ and }
\cot{\theta_H} = \frac{m_{hot}}{e B \tau_{hot}}\frac{2r}{(1+r)} \ ,
\label{eq1}
\end{equation}
where $r = l_{hot}/l_{cold}$ is the ratio of the mean free paths along the hot spots  and the cold spots  directions, $R_H^\infty$ is the high-temperature asymptotic value of $R_H$, $m_{hot}$ and $\tau_{hot}$ are respectively the effective mass and scattering time in the strong scattering directions. 
This description is model dependent only in the choice of the anisotropy of the scattering. It thus allows the determination of the scattering along the hot and cold directions, $m_{hot}/\tau_{hot}$ and $m_{cold}/\tau_{cold}$. 
As underlined in Ref.~\onlinecite{stojkovic97}, the results do not strongly depend on the exact value for $R_H^\infty$. In our case, the  determination of the latter  is complicated by the linear dependence for $R_H(T)$, with no observable saturation at high temperature. We used an estimate for $R_H^\infty$ for the most OD state where the temperature dependence of the coefficient is the weaker and assumed that $R_H^\infty$ is proportional to the maximum in $R_H(T)$. We checked that the results were essentially independent from the choice of $R_H^\infty$ for the most OD state. We also evaluated $R_H^\infty$ from the hole content, as estimated from the superconducting transition temperature and its phenomenological relation to the carrier concentration\cite{takagi89,presland91}, which yielded results similar to the ones obtained from the first method, presented in Fig.~\ref{scattering} (using $R_H^\infty = 7\,10^{-10}$ m$^3$ C$^{-1}$ for the higher doping).

\begin{figure}

\includegraphics[width= \columnwidth]{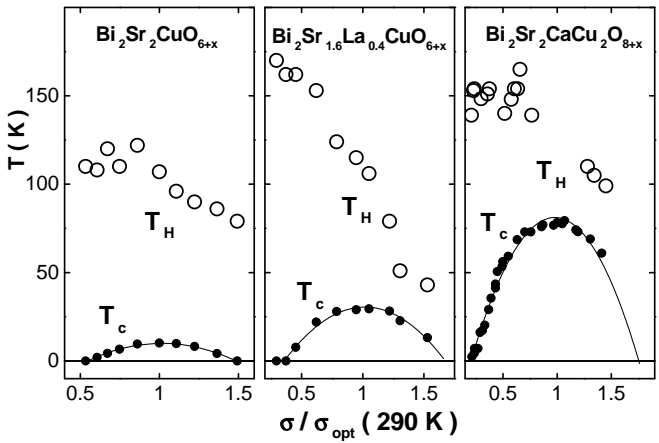}
\caption{Temperature at which $R_H(T)$ deviates from a high temperature linear fit by $\Delta\,R_H(T_H) = -5\,10^{-11}$~m$^3$/C. Full lines are parabolic fits of $T_c$. Data for La-Bi2201 and Bi2212 are from Refs.~\onlinecite{konstantinovic2000} and \onlinecite{konstantinovic2000b}.}\label{deviation}
\end{figure}

The $T$ dependence of the hot-spot scattering rates is markedly different for the OD and UD samples. To emphasize this, we normalized the curves by their room temperature value (Fig.~\ref{scattering}). Doing so, it is apparent that these curves split into two groups: the scattering rate at the hot spots for the OD states  merge onto a single curve which shows no saturation at 300~K, while it is strongly curved and clearly  shows a saturation for the UD ones. This behavior is qualitatively what is expected from the nearly-antiferromagnetic Fermi-liquid model, where spin fluctuations enhance the scattering at the intersection of the Fermi surface and the Brillouin zone along \textbf{k}=$\pm(0,\pi)$\cite{montoux91}. Within this framework, the non saturating behavior and the saturating one for the hot-spots scattering rate correspond to the mean-field, OD and to the pseudogap, UD regimes respectively. It is however not clear why the scattering rates  should all merge onto a single curve in the OD regime: this seems to indicate that, above optimal doping, the mean-field regime is valid over the entire temperature range, with no crossover to the pseudogap one at low $T$, as would be expected in the case of a steep boundary between the metallic phase and the pseudogap one as one crosses optimal doping\cite{yurgens2003}.

Nevertheless, this approach allows to clearly distinguish between the UD and OD regimes in the present case. It is remarkable that this phenomenological description is able to account for a maximum in $R_H(T)$. Such a behavior corresponds to an offset in the hot and cold spots scattering rates that accounts for a finite mean free path in the limit of zero temperature, as is clearly seen in Fig.~\ref{scattering}. This allows for a decreasing anisotropy 
with a finite limit,
as $T \rightarrow 0$. This is the reason why data in Ref.~\onlinecite{ito93} for YBa$_2$Cu$_3$O$_{6.68}$, which show a maximum in $R_H(T)$, also yield a scattering rate with a finite value in the limit $T \rightarrow 0$\cite{stojkovic97}. This may be easily checked by adding to the phenomenological scattering rates in ref.\onlinecite{stojkovic97} a 
T-independent contribution, and inverting using Eq. \ref{eq1}. The effect of this additional impurity contribution to the scattering rate was explicitely considered in ref.~\onlinecite{hussey2003}. It was shown that a decrease of the p value from the intrinsic $p=2$ value, as observed here, can be accounted by a decrease of the scattering rate anisotropy. In the present case, the disorder at the origin of the T-independent scattering may originate from the strong modulation, or from both in-plane and out-of-plane impurities.

There are several other models that aim to account for the temperature dependence of the Hall coefficient. Hussey's model accounts for a saturation of the hot spots scattering rate for the OD regime with a lower cutoff of the mean path. Another Fermi liquid theory is the one proposed by Bok \etal, using hole-like and electron-like orbits, depending on the carrier energy with respect to the one of a Van Hove singularity\cite{bok2004}. This model, as well as the one by Stojkovic \etal, have been criticized  in ref.~\onlinecite{castro2004}. However, Castro \etal{} recognize that their scattering times are qualitatively similar to the ones in Ref.~\onlinecite{stojkovic97}. Recently also, the marginal Fermi liquid theory was shown to provide transport coefficients in agreement with experiments [including the sub linear $\cot(\theta_H)(T^2)$], when the anisotropic $T$-independent scattering rate is adequately chosen\cite{abrahams2003}. However, it is difficult to understand, within such a model, how the scattering rate computed from our data could be insensitive to the Fermi liquid to the marginal Fermi liquid crossover, when going from the OD regime to the optimally doped one.

Besides these approaches, there have been attempts to evaluate the importance of superconducting fluctuations either directly from the temperature dependence of the Hall coefficient or from the Hall conductivity. Indeed, it is known that superconducting fluctuations bring a contribution to the Hall conductivity\cite{fukuyama71}. Rice \etal{} have analyzed in this way the Hall conductivity in YBa$_2$Cu$_3$O$_{7-x}$\cite{rice91}. Further, Matthey \etal{} suggest that unconventional fluctuations may be uncovered by the low temperature deviation of $R_H$ from a $1/T$ behavior\cite{matthey2001} and the crossover to the pseudogap regime was shown to directly affect the Hall constant\cite{ito93,hwang94}. We have plotted in Fig.~\ref{deviation} the temperature $T_H$ at which $R_H$ deviates by a given quantity from a linear high temperature fit obtained in the range 200--300~K, as well as for La-doped Bi2201 and Bi2212 from earlier results\cite{konstantinovic2000,konstantinovic2000b}. Such a fit does not preclude the possibility that $R_H$(T) may be described as by  an  hyperbolic function  on a larger range of $T$, but this provides a simple way to estimate this deviation and was found equally convenient in the case of the other compounds of the Bi-based family. Clearly, Bi-2201 stands aside from the two other compounds: while $T_H$ appears to merge with the superconducting transition temperature on the OD side of the phase diagram for both Bi,La-2201 and Bi-2212, a large temperature interval remains between $T_c$ and $T_H$ for the most OD state in the case of Bi-2201:  $T_H$ is only weakly dependent on doping. This observation strikingly correlates with the previous one, that the exponent for $\cot(\theta_H)$ is also only weakly dependent on doping. It contradicts Matthey \etal{} proposal that the Hall data for several cuprates supports the existence of some intermediate crossover temperature between $T_c$ and  the pseudogap temperature\cite{matthey2001}, where Cooper pairs start to form locally\cite{devillard2000,timm2002,curty2003}, which could be detected from the maximum in $R_H$(T), laying in our case close to the deviation point at $T_H$. In the present case, this would imply that there is a large temperature domain above $T_c$, \textit{only weakly dependent upon the doping range}, where these incoherent Cooper pairs are present. This is problematic if the presence of such Cooper pairs is  only expected in  the pseudogap state, in the UD regime. 

So, the following general picture may be proposed. Strongly disordered materials, as is the case here, can display a maximum in $R_H(T)$ well above $T_c$, independently from the doping state, simply due to a relatively small mean free path in the limit $T \rightarrow 0$. In this case, the balance between the anisotropic 
T-dependent
scattering mechanism and
a T-independent impurity one sets the temperature for the maximum. 
We note that, for weakly disordered materials, superconducting fluctuations account for an increasing $R_H$ with temperature, close to $T_c$, on an otherwise decreasing function of temperature as obtained from anisotropic scattering mechanism as in Ref.~\onlinecite{stojkovic97}. When doping is decreased, the carriers tend to localize at low $T$, and the deviation of $R_H(T)$ from its high temperature behavior provides a temperature that shifts above the superconducting one, with no need of the intervention of additional superconducting fluctuations.

\end{document}